# THE RADIO SPECTRUM OF SGR A*

WOLFGANG J. DUSCHL[1,2] and HARALD LESCH[1]
1: *Max-Planck-Institut für Radioastronomie*
   *Auf dem Hügel 69, D-53121 Bonn, Germany*
2: *Institut für Theoretische Astrophysik*
   *Im Neuenheimer Feld 561, D-69120 Heidelberg, Germany*

**Abstract.** We discuss the radio spectrum of Sgr A* in the frequency range between $\approx 1\,\mathrm{GHz}$ and $\approx 1\,000\,\mathrm{GHz}$, show that it can be explained by optically thin synchrotron radiation of relativistic electrons, and point toward a possible correlation between the spectrum of Sgr A* and larger-scale ($\lesssim 50\,\mathrm{pc}$) radio emission from the Galactic Center region.

## 1. Introduction

Duschl and Lesch (1994 = DL94) have shown that one can understand the radio spectrum of Sgr A* between $\approx 1$ and $\approx 1\,000\,\mathrm{GHz}$ as being due to a single physical process, namely optically thin synchrotron radiation of relativistic electrons. In this contribution, we will shortly discuss their results (sect. 2), especially in the light of new observational results (Zylka *et al.*, 1995), and compare it with other models for Sgr A* (sect. 3). Finally, we will comment on the large-scale radio emission of the Galactic Center and its implications for the radiation mechanism in the Galactic Center (sect. 4).

## 2. Optically Thin Synchrotron Radiation of Relativistic Electrons

For the frequency range from $\approx 1\,\mathrm{GHz}$ up to a few hundred GHz, the time averaged flux density $F_\nu$ shows a dependency on frequency $\nu$ of $F_\nu \propto \nu^{1/3}$ (fig. 1; for details, and for a compilation of the then available observations in this frequency range, see DL94). While MIR observations ($8\,\mu\mathrm{m} \leq \lambda \leq 20\,\mu\mathrm{m}$) give only upper limits for the flux densities, nonetheless they show that the radio spectrum must attain its maximum somewhere around $10^3$ Ghz with a sharp drop of the flux density toward higher frequencies. DL94 have demonstrated that one can interpret the spectrum of Sgr A* in the radio to submillimeter range as due to a single physical process: optically thin synchrotron emission of relativistic electrons. In fig. 1 we show three model spectra for the extremal case of monoenergetic electrons (labelled a, b, and c). For the flux density $F_\nu$ we get in this case

$$F_\nu \propto \nu^{1/3} \exp\left(-\nu/\nu_c\right)$$

with a cut-off frequency $\nu_c$. The physical parameters for the three model spectra are summarized in table 1. We find that we need typically magnetic fields of $\approx 5$ G and electron energies $\approx 120$ MeV, only weakly dependent of the frequency $\nu_{\mathrm{max}}$ where the flux density attains its maximum. Values for $\nu_{\mathrm{max}}$ considerably lower than $\approx 500$ GHz are excluded by the measured flux densities, while $\nu_{\mathrm{max}} \gtrsim 1500$ GHz would lead to contradictions with the upper limits of the MIR observations.

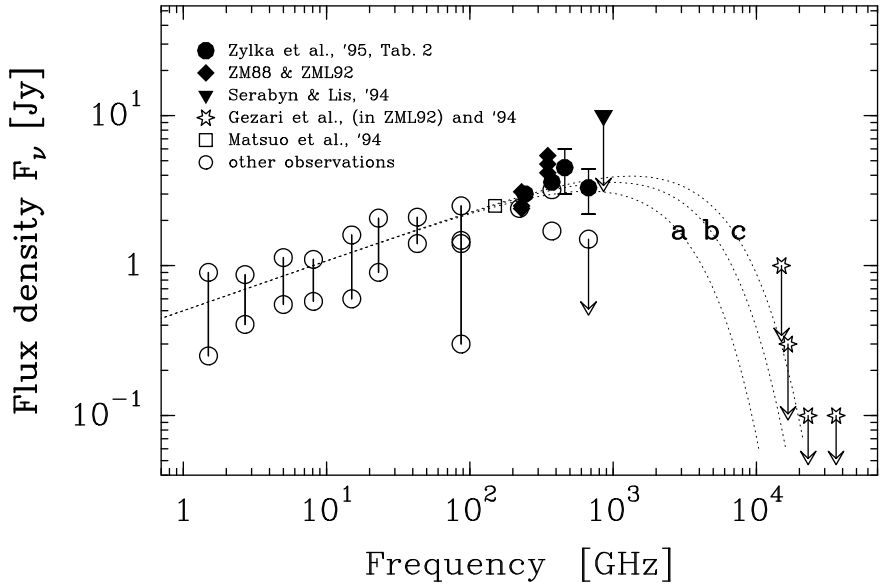

*Figure 1.* The radio spectrum of Sgr A*. Vertical arrows indicate upper limits; vertical lines around individual symbol indicate error bars; vertical lines connecting symbols indicate observed variability ranges [ZM88: Zylka and Mezger (1988); ZML92: Zylka, Mezger, and Lesch (1992); for details on "other observations", see DL94]. The dotted lines are optically thin synchrotron model spectra for three different sets of physical parameters (a, b, and c; see table 1). [Figure after Zylka *et al.*, 1995]

TABLE 1. Physical parameters for the synchrotron emission models of Sgr A* as shown in fig. 1. [$\nu_c$: cut-off frequency; $\nu_{max}$: frequency of maximum flux density; $B$: magnetic field strength; $E$: energy of (monoenergetic) electrons; $L$: resulting radio luminosity of Sgr A*]

| Model | $\nu_c$ | $\nu_{max}$ | $B$ | $E$ | $L$ |
|---|---|---|---|---|---|
| | [GHz] | [GHz] | [G] | [MeV] | [$L_\odot$] |
| a | $2\,10^3$ | $6.7\,10^2$ | 5.1 | 111 | $2.5\,10^2$ |
| b | $3\,10^3$ | $1.0\,10^3$ | 5.6 | 121 | $4.3\,10^2$ |
| c | $4\,10^3$ | $1.3\,10^3$ | 6.0 | 127 | $6.3\,10^2$ |

The exact values for, e.g., the magnetic field strength and the electron energy depend on the volume from which the emission comes; here we have taken a spherical volume with a radius of $4\,10^{13}$ cm, according to Krichbaum *et al.*'s (1993) VLBI resolution of Sgr A*. A magnetic field strength of the required order of magnitude can easily be achieved in an accretion disk around a black hole of a few $10^6\,M_\odot$ and an accretion rate of $\approx 10^{-7\ldots-6}\,M_\odot/$yr, i.e., a situation presumably typical for the immediate vicinity of Sgr A* (Falcke *et al.*, 1993a). Recent observations at $\lambda\,1\,300, 800, 600$, and $450\,\mu$m (Zylka *et al.*, 1995), and at $\lambda\,2$ mm (Matsuo *et al.*, 1994) are in good agreement with this picture and help to constrain the frequency where the spectrum attains its maximum to $\nu_{max} \in [700\,\text{GHz}, 1\,200\,\text{GHz}]$.

While in the lower frequency range ($\nu \lesssim 100$ GHz) Sgr A* clearly shows variability over time scales ranging down to weeks (Zhao *et al.*, 1992), at higher frequencies the situation is less clear yet. If any, only the variability at $\nu \lesssim 10$ GHz is attributed to interstellar szintillation. Thus, at least the variability at frequencies $\gtrsim 10$ GHz most likely is due to Sgr A* and/or its immediate vicinity. Interestingly enough, the time scale of the viscous evolution of the inner radii of an accretion disk around a $10^6\,M_\odot$ black hole at an accretion rate of around $10^{-6}\,M_\odot/$yr is of the same order as that of the observed variability of Sgr A*. However, our new observations show, at best, marginal evidence for variability at $\lambda\,1\,300\,\mu$m within a time scale of ten days. This clearly warrants a longer term monitoring of the spectrum of Sgr A* in this spectral range.

It is important to note that for the applicability of our model a strictly monoenergetic energy distribution for the electrons is not required (Crusius and Schlickeiser, 1988; Beckert *et al.*, 1995).

## 3. Other Models for Sgr A*

Falcke *et al.* (1993b; also this volume) interprete the radio emission as due to a jet, which has been observed at $\lambda 7$ mm by Krichbaum *et al.* (1993). In their model the observed synchrotron radiation is the result of an evolving electron spectrum injected into a supersonic, freely expanding conical jet, convecting a tangled magnetic field. The electrons, accelerated via shock waves in the jet, follow a power law energy distribution with $n(\gamma) \propto \gamma^{-2}$, with a low and a high energy cut-off ($\gamma$: Lorentz factor; $n$: volume density of electrons). They obtain a flat spectrum by integrating over optically thin and thick jet components.

Melia (1994) explains the radio spectrum of Sgr A* as due to magnetic bremsstrahlung of a plasma with temperatures of $\sim 10^{10}$ K descending toward the horizon with a rate of $\sim 1.5\, 10^{-4} M_\odot/$yr. According to his figure 8, the observed radio spectrum would be produced by very hot electrons in a magnetic field of far more than $1\,000$ G, in order to produce a break frequency of $\sim 1\,000$ GHz.

Whereas the two above mentioned models also favor a scenario with a black hole mass of $\sim 10^6 M_\odot$, Mastichiadis and Ozernoy (1994) obtain an upper limit for the black hole mass of $\sim 3\, 10^3 M_\odot$ by considering X-ray and $\gamma$-ray emission from the Galactic Center region. However, concerning the radio emission their model seems to us to be not applicable. Their estimated magnetic field due to spherical accretion from the surrounding stars is $\sim 4\, 10^4$ G at the Schwarzschild radius. Electrons emitting in such a strong magnetic field must have Lorentz factors of the order unity, which means they radiate in a cyclotron line but not in a synchrotron continuum. Furthermore, they give a minimum Lorentz factor for the electrons $\sim 10^4$. These electrons would only allow for a field strength of less than $10^{-3}$ G to emit at 100 GHz. For their spherical model such a field strength is obtained at $\sim 4\, 10^{14}$ cm, i.e., at $10^6$ Schwarzschild radii for a $10^3 M_\odot$ black hole. But the observations at $\lambda 7$ and 3 mm (Krichbaum *et al.*, 1993, 1994) already resolved source sizes of $\sim 0.13 - 0.33$ mas ($1.6 - 4\, 10^{13}$ cm).

Summarizing, any synchrotron interpretation of the radio spectrum of Sgr A* sets a constraint on its size. The observed luminosity is proportional to the volume and to the square of the magnetic field strength. If one favors smaller black hole masses, the volume decreases, whereas the field strength increases. To emit still at the same frequency, the particles' energy has to decrease. However, a particle with a Lorentz factor of around unity radiates cyclotron line emission instead of a synchrotron contiuum spectrum.

## 4. The large-scale radio emission of the Galactic Center

The large-scale ($\simeq 50\,\mathrm{pc}$) radio emission of the galactic center region was observed by Reich *et al.* (1988). These observations include the very center Sgr A* and the extended components up to a distance of 40 pc, often referred to as "Bridge" and "Arc" (Brown and Liszt, 1984), located north of Sgr A*. They measured a similarly flat or inverted spectrum in the frequency range betwen 843 MHz and 43 GHz in the Bridge and Arc as in Sgr A*. After a separation of thermal/nonthermal emission they find, that most of the nonthermal emission has an inverted spectrum with $\alpha \simeq 0.3$ ($\alpha = d\log F_\nu / d\log \nu$). Whereas the spectrum of the Arc can be explained by local acceleration (Lesch and Reich, 1992), the spectral index $\alpha \sim 0.3$ in the Bridge strongly suggests a very close physical connection between the extended emission and the radio source Sgr A*, which was investigated by Lesch *et al.* (1988).

If Sgr A* is the source of the electrons which radiate in the Bridge – as may be indicated by the common spectral index – this has important implications. It makes it highly unlikely that the emission of Sgr A* is due to superposed optically thick synchrotron components. This can be seen as follows: let us assume the emission of Sgr A* is composed of optically thick synchrotron sources. When they move into the extended regions, they become optically thin. At frequencies higher than the turnover frequency, which defines the transition from optically thin to optically thick, the spectrum is a power law with negative spectral index. Such a behaviour is well known from the time evolution of synchrotron components in active galactic nuclei (see, e.g., Marscher and Gear, 1985). The finding that $\alpha$ is not only positive but even still 0.3 contradicts the assumption that the emission is optically thick synchrotron radiation.